\documentclass{article}  
\usepackage{spadre2008}
\usepackage{graphicx}
\frompage{000} \topage{000}                                              
\def\rightmark{Heavy Flavor in the sQGP}
\def\leftmark{R.Rapp et al.}
 
\title{Heavy Flavor in the sQGP} 
\authors{
{R. Rapp$^{1}$, D. Cabrera$^2$, V. Greco$^3$, M. Mannarelli$^4$, 
H. van Hees$^5$ %
}\\[2.812mm]
{\normalsize
\hspace*{-8pt}$^{1}$ Cyclotron Institute and Physics Department, 
Texas A\&M University, \\College Station, TX 77843-3366, USA\\[0.2ex] 
\hspace*{-8pt}$^{2}$ Departamento de F\'isica Te\'orica II, Universidad 
Complutense, 28040 Madrid, Spain\\[0.2ex]
\hspace*{-8pt}$^{3}$ 
INFN-LNS, Via S. Sofia 64, I-95125 Catania, Italy\\[0.2ex]
\hspace*{-8pt}$^{4}$ Instituto de Ciencias del Espacio (IEEC/CSIC),
Facultat de Ci\`encies, \\ Torre C5, E-08193 Bellaterra (Barcelona), 
Spain\\[0.2ex]
\hspace*{-8pt}$^{5}$ Institut f\"ur Theoretische Physik,
Justus-Liebig-Universit\"at Giessen, \\
D-35392 Giessen, Germany\\[0.2ex]
}}
 
\abstract{We attempt a unified treatment of heavy quarkonia
and heavy-quark diffusion in the Quark-Gluon Plasma. Our approach is
based on finite-temperature $T$-matrices with interaction potentials 
estimated from the heavy-quark internal energy computed in thermal 
lattice QCD (lQCD). In the charmonium sector $S$-wave bound 
states ($J/\psi$, $\eta_c$) survive up to temperatures of $\sim$2$T_c$, 
not inconsistent with constraints from euclidean correlation functions
in lQCD.  In the open-heavy flavor sector, the $T$-matrix interaction 
reduces heavy-quark diffusion substantially, leading to fair agreement 
with single-electron spectra at RHIC and suggestive for a small
viscosity-to-entropy ratio close to $T_c$.  
  }

\keyword{Quark-Gluon Plasma, Heavy Quarks, Heavy-Ion Collisions} 
\PACS{25.75.Nq, 24.10.Cn, 14.65.Dw}
 
\begin{document}
 
\maketitle
\setcounter{page}{1}

\section{Introduction}
\label{sec_intro}
The properties of the strong-interaction matter produced in Au-Au
collisions at RHIC are under intense debate. The empirical evidence
collected thus far suggests a liquid-like medium which 
(i) quickly establishes, and then maintains, local thermal equilibrium 
for low transverse-momentum particles ($p_t$$\le$2~GeV),
(ii) is largely opaque to high $p_t$$\ge$5~GeV particles, 
(iii) exhibits constituent quark-scaling properties in hadron production 
at intermediate $p_t$$\simeq$2-5~GeV. A key question is if 
and how these phenomena are related~\cite{Rapp:2008qc,Majumder:2007zh} 
and what the relevant interactions are (e.g., radiative or elastic, 
perturbative or nonperturbative, gluon- or quark-driven, etc.).  
In this article we approach this problem via the heavy-quark (HQ) sector.
For heavy quarkonia, the hope has emerged that, using potential models, 
one can quantitatively analyze in-medium quarkonium properties based
on input from lattice QCD (lQCD), which is discussed in Sec.~\ref{sec_onia}. 
The question arises whether similar interactions could be relevant 
for low-momentum properties of individual heavy quarks in the Quark-Gluon
Plasma (QGP), i.e., for HQ diffusion; this is studied in Sec.~\ref{sec_open},
including applications to observables. We conclude in Sec.~\ref{sec_concl}.

\section{Heavy Quarkonia}
\label{sec_onia}  
Our starting point is a two-body scattering equation for the in-medium
$T$-matrix of a heavy quark ($Q$) and antiquark 
($\bar Q$)~\cite{Mannarelli:2005pz,Cabrera:2006wh},
\begin{equation}
T(E,q,q') = V(q,q') + \int d^3k \ V(q,k) \ G(E,k) \ T(E,k,q') \ .
\label{tmat}
\end{equation}
It represents a ``ladder'' resummation of a suitably defined HQ potential,
$V$, and is widely used in the nuclear many-body problem as well as in 
the study of electromagnetic plasmas~\cite{Ebeling:1986}; 
$G$ denotes the intermediate 
$Q$-$\bar Q$ propagator including selfenergy insertions due to
interactions with the surrounding medium particles. The main 
approximations are neglecting (i) virtual $Q$-$\bar Q$ excitations, and
(ii) retardation effects in the interaction; both should be 
reasonable for large quark masses, $m_Q$. The $T$-matrix provides a
uniform treatment of bound and scattering states which is particularly
important for situations involving dissolving bound states, such
as for atoms in plasma physics or quarkonia in the QGP. 
\begin{figure}[!t]
\includegraphics[width=0.45\textwidth]{pot-CR06.eps}
\hspace{0.3cm}
\includegraphics[width=0.47\textwidth]{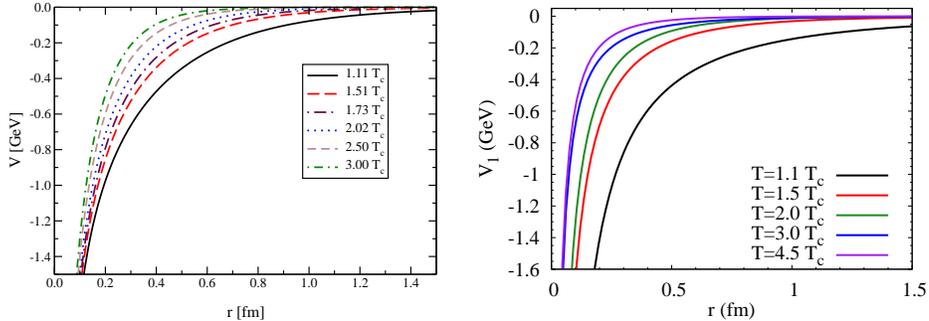}
\vspace{-0.4cm}
\caption{Color-singlet heavy-quark internal energies at finite
temperature (with the asymptotic value, 
$U_\infty$$\equiv$$U(r\to\infty,T)$ subtracted) 
extracted from 3-flavor~\cite{Petreczky:2004pz} (left 
panel)~\cite{Cabrera:2006wh} and quenched lQCD~\cite{Kaczmarek:2003dp}
(right panel)~\cite{Wong:2004zr}.}
\label{fig_pot}
\end{figure}
The precise relation between the HQ potential and free energy, 
$F=U-TS$, computed in thermal lQCD, is an open problem.
We here adopt the internal energy, $U$, as the potential,
but other choices are possible~\cite{Mocsy:2007yj,Alberico:2007rg}. 
Figure~\ref{fig_pot} shows that different extractions of $U$ still
imply up to $\sim$40\% uncertainty.
In either case, the increased color-screening leads to a reduction in 
charmonium binding with increasing temperature. However, when including 
an in-medium reduction of the $c$-quark mass governed by the asymptotic 
value of the internal energy, $m_c^*$=$m_c^0$+$U_\infty/2$,
the bound-state mass is almost stable. 
\begin{figure}[!t]
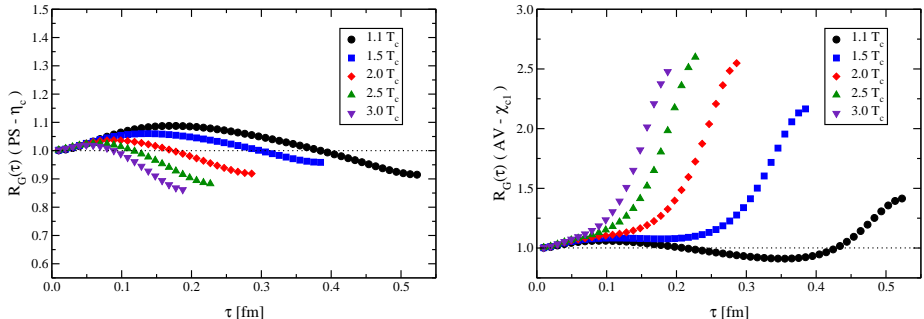

\includegraphics[width=0.45\textwidth]{RGccPS-tmat.eps}
\hspace{0.5cm}
\includegraphics[width=0.45\textwidth]{RGccAV-tmat.eps}
\vspace{-0.4cm}
\caption{Euclidean correlators for $\eta_c$ (left) and $\chi_{c1}$
(right) in the QGP computed within 
the $T$-matrix approach~\cite{Cabrera:2006wh} using the HQ internal 
energy extracted from $N_f$=3 lQCD~\cite{Petreczky:2004pz}. 
The correlators are normalized to a reconstructed one based on the 
vacuum $T$-matrix. The increase in the $\chi_c$ correlator is largely
driven by zero-mode contributions~\cite{Umeda:2007hy} implemented
in quasi-particle approximation.} 
\label{fig_RGtmat}
\end{figure}
Together with the lowering of the $c\bar c$ threshold, 
$E_{\rm thr}$=2$m_c^*$, and nonperturbative rescattering strength 
generated by the $T$-matrix~\cite{Cabrera:2006wh}, the pertinent 
$\eta_c$ correlator is rather stable (cf.~left panel of 
Fig.~\ref{fig_RGtmat}), similar to the findings in 
lQCD~\cite{Datta:2003ww,Morrin:2005zq}. The underlying spectral 
functions show that the bound state ``melts'' slightly above 
$\sim$2$T_c$. Smaller melting temperatures are found in
Refs.~\cite{Mocsy:2007yj,Alberico:2007rg} based on different input
potentials, but the lQCD correlators can be well reproduced,
due to the above-mentioned interplay between binding and 
threshold effects.
An independent determination of the in-medium HQ mass, an
improved definition of the potential, a quantitative implementation 
of finite width effects, as well as a coupled channel treatment 
to account for gluonic excitations, may be required to make 
further progress. 

\section{Heavy-Flavor Transport and Observables}
\label{sec_open}
In Ref.~\cite{vanHees:2004gq} the exchange of effective meson resonances 
in scattering of heavy quarks in the QGP has been introduced and found 
to reduce charm- and bottom-quark thermalization times by a factor of 
$\sim$3 compared to elastic perturbative QCD (pQCD) scattering. When 
implemented into relativistic Langevin simulations for Au-Au collisions 
at RHIC~\cite{vanHees:2005wb}, the 
predictions for semileptonic electron ($e^\pm$) spectra from HQ decays 
turned out to be in fair agreement with 
experiment~\cite{Adare:2006nq,Abelev:2006db}.  
\begin{figure}[!t]
\includegraphics[width=0.47\textwidth]{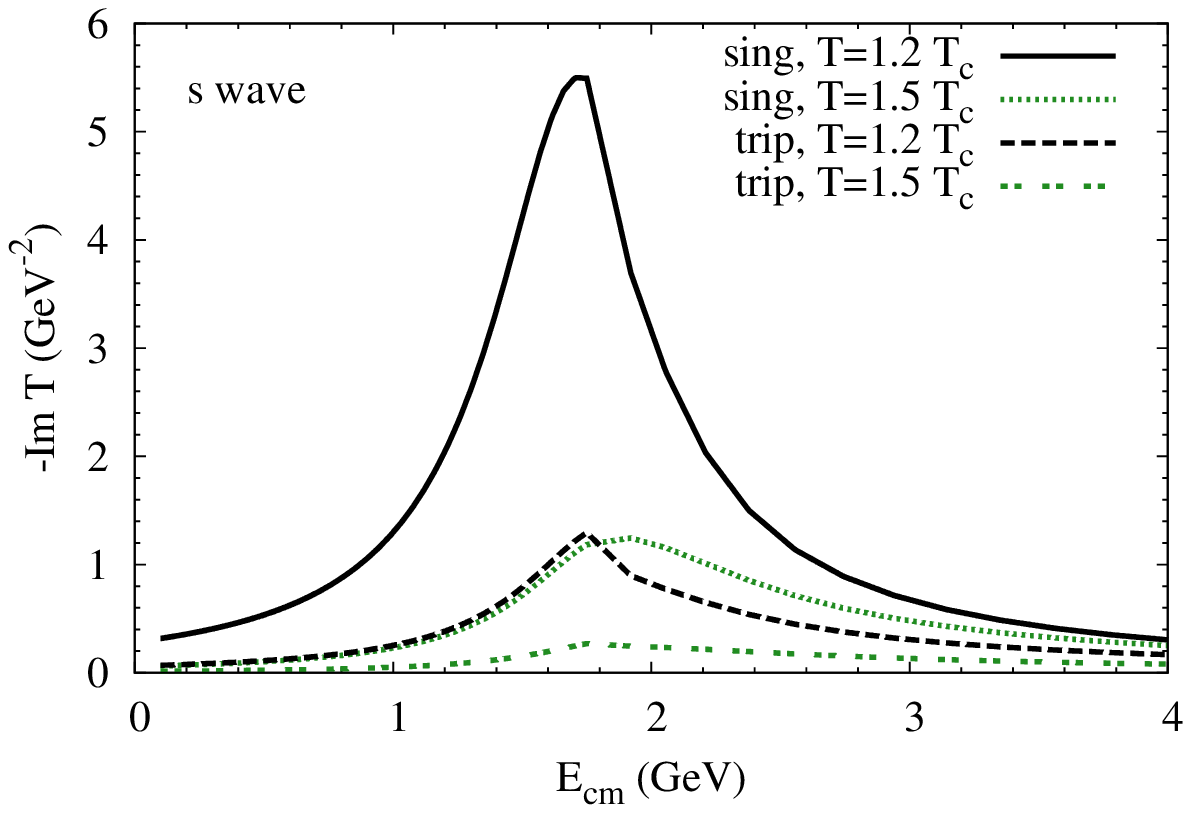}
\hspace{0.3cm}
\includegraphics[width=0.45\textwidth]{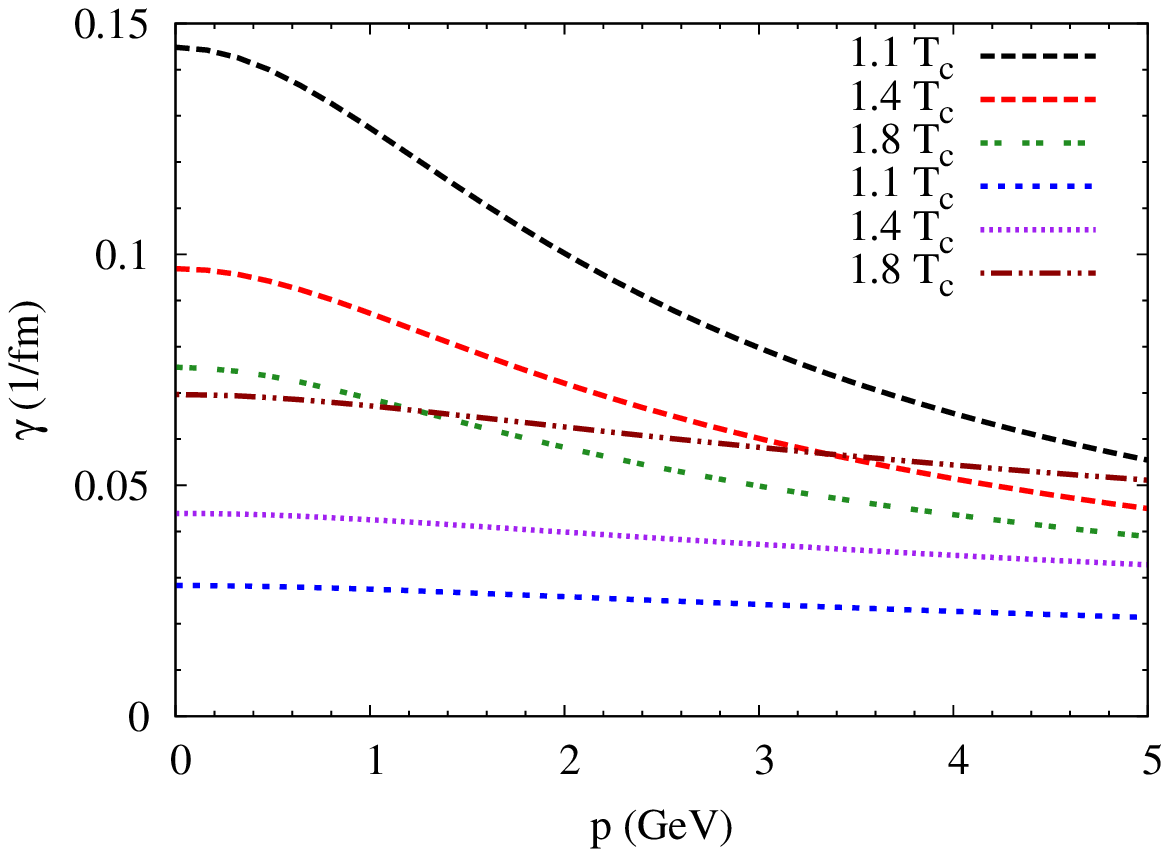}
\vspace{-0.4cm}
\caption{Left panel: Imaginary part of the $T$-matrix for $c$-quark scattering
of light quarks and antiquarks in the QGP
Right panel: thermal relaxation rates for $c$-quarks following from
the nonperturbative heavy-light $T$ matrices (upper curves at $p$=0)
and from LO pQCD scattering off quarks and gluons with $\alpha_s$=0.4.}
\label{fig_tmat}
\end{figure}
In Ref.~\cite{vanHees:2007me}, nonperturbative HQ interactions in the
QGP have been evaluated by employing the same $T$-matrix approach as
discussed in the previous section, including all color channels (1,
$\bar 3$, 6, 8) as well as $S$- and $P$-waves in heavy-light quark
scattering.  The pertinent $T$-matrices exhibit (``pre-hadronic'')
resonance-like $D$-meson and diquark correlations up to temperatures of
$\sim$1.5 and 1.2~$T_c$, respectively, cf.~left panel of
Fig.~\ref{fig_tmat} (using the potential in the right panel of
Fig.~\ref{fig_pot}; in the bottom sector, the dissolution temperatures
are slightly larger). The repulsive sextet and octet channels, as well
as $P$-waves, are suppressed. The dissolution of the resonances leads to
a {\em decrease} of the thermalization rate with increasing temperature,
opposite to the standard behavior as found, e.g., in pQCD, cf.~right
panel of Fig.~\ref{fig_tmat}.

The heavy-light quark $T$ matrices (supplemented by pQCD interactions 
with gluons) have been used to compute HQ diffusion in a Fokker-Planck 
approach and applied to Au-Au collisions at RHIC utilizing relativistic 
Langevin simulations in an expanding fireball. 
\begin{figure}[!t]
\begin{center}
\includegraphics[width=0.46\textwidth]{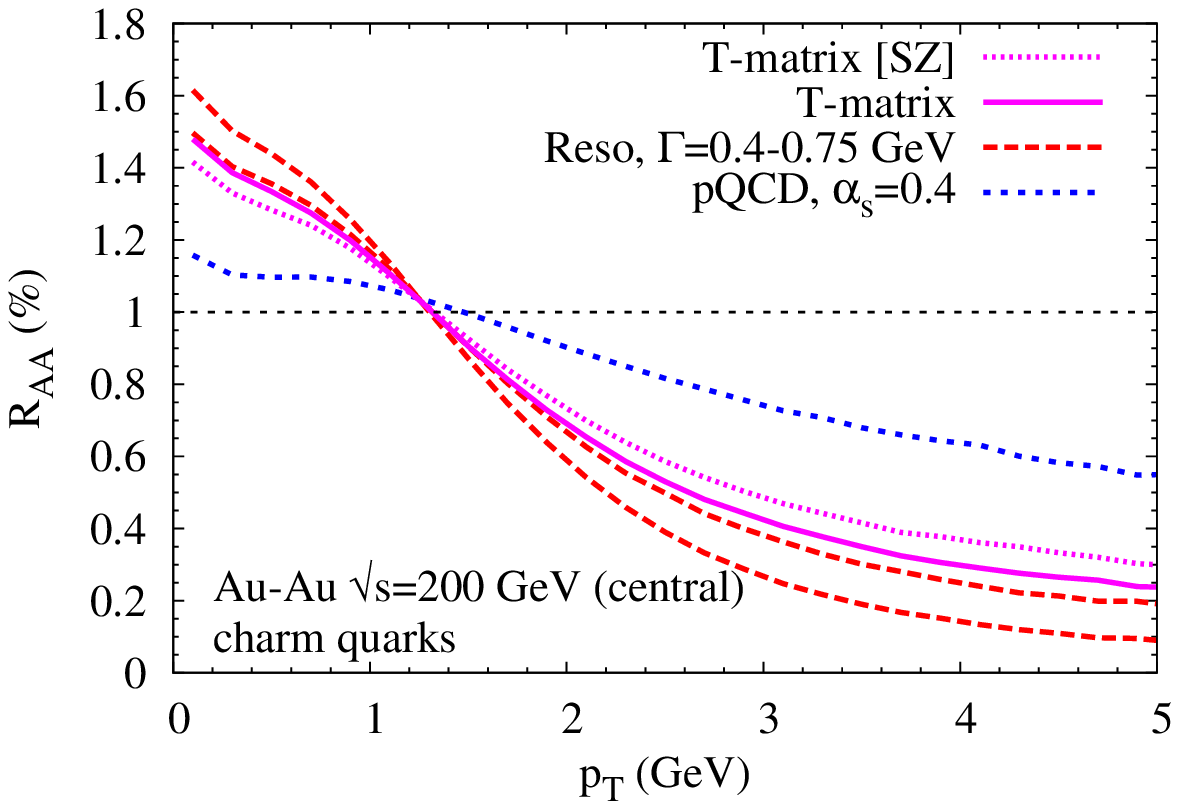}
\hspace{0.2cm}
\includegraphics[width=0.47\textwidth]{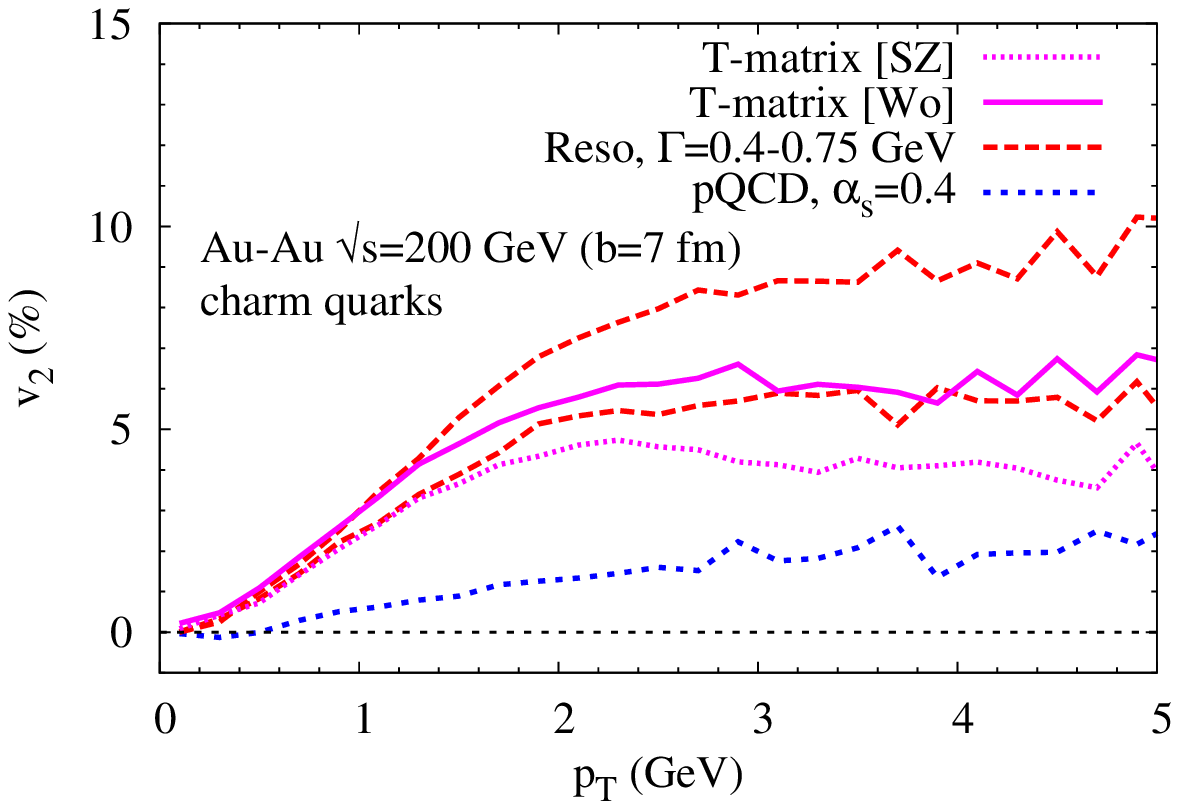}
\includegraphics[width=0.46\textwidth]{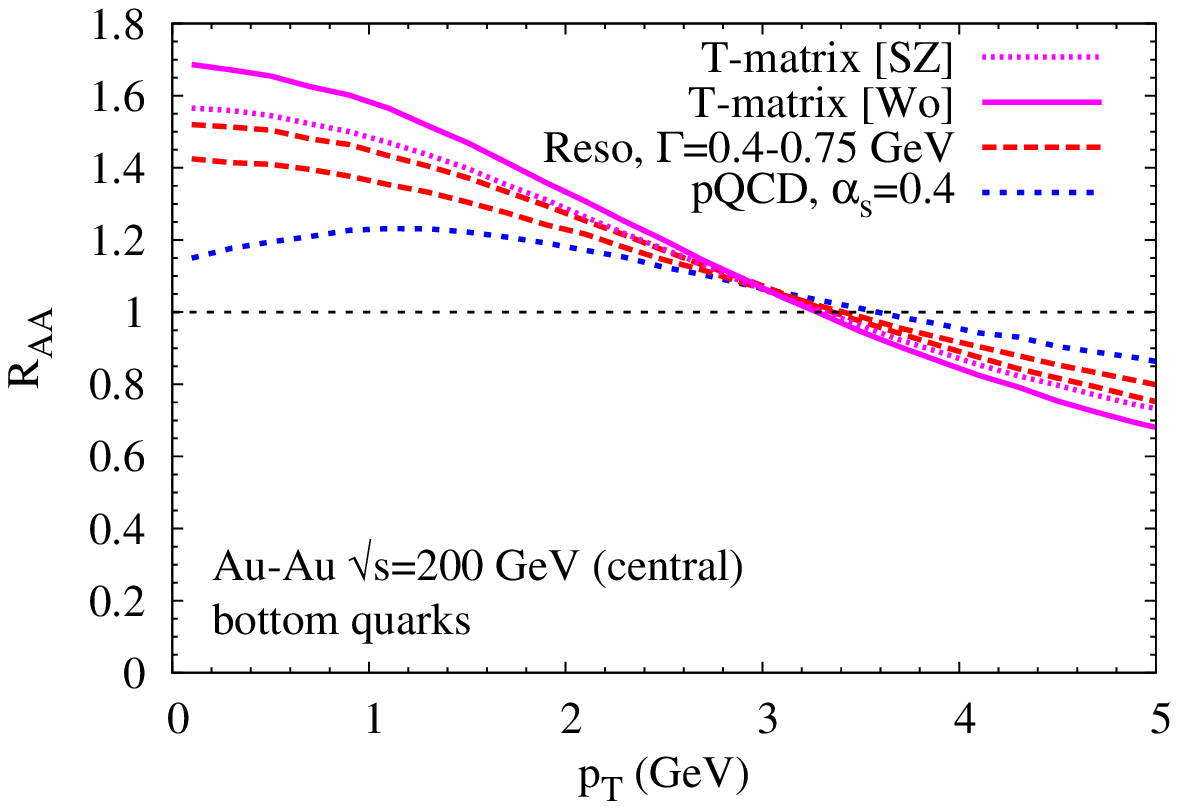}
\hspace{0.3cm}
\includegraphics[width=0.46\textwidth]{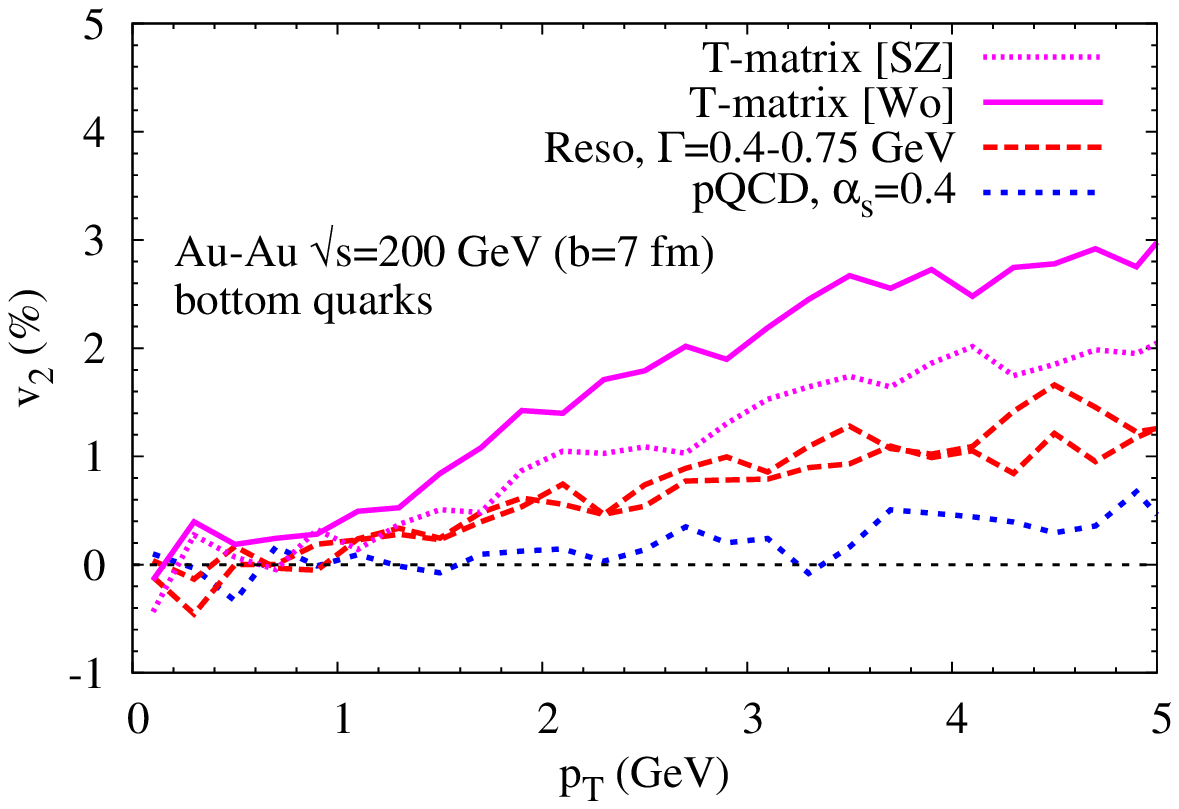}
\end{center}
\vspace{-0.6cm}
\caption{HQ spectra in Au-Au collisions at RHIC utilizing relativistic
Langevin simulations with nonperturbative HQ
interactions~\cite{vanHees:2005wb,vanHees:2007me}.}
\label{fig_hq-spec}
\end{figure}
The resulting nuclear modification factor and elliptic flow are quite 
comparable to the effective resonance 
model~\cite{vanHees:2004gq,vanHees:2005wb} (see Fig.~\ref{fig_hq-spec}), 
with slightly smaller (larger) effects for $c$ ($b$) quarks (which is 
qualitatively similar to the dissociation model of Ref.~\cite{Adil:2006ra}). 
Uncertainties due to different extractions of the lQCD internal 
energy~\cite{Wong:2004zr,Shuryak:2004tx} amount to $\sim$30\%. The HQ 
spectra have been hadronized into $D$ and $B$ mesons in a combined 
coalescence/fragmentation framework~\cite{Greco:2003vf,vanHees:2005wb},
with subsequent semileptonic (3-body) decays into single electrons.
The relative weight of charm and bottom contributions is estimated from 
$d$-Au data~\cite{star-dAu}, crossing at about $p_t^e$$\simeq$5~GeV,
(cf.~also Ref.~\cite{Awes:2008qi}).
Note that quark coalescence at $T_c$ naturally emerges from the 
``pre-hadronic'' resonant correlations in the $T$-matrix~\cite{Ravagli:2007xx}.
The electron spectra in Au-Au collisions~\cite{vanHees:2007me}, which do
not involve adjustable parameters, compare well to the most recent RHIC
data~\cite{Adare:2006nq,Abelev:2006db,Awes:2008qi}, see left panel of
Fig.~\ref{fig_elec}. The HQ diffusion coefficient at $p$=0 can be used
for a schematic estimate of the viscosity-to-entropy ratio ($\eta/s$) 
in the QGP~\cite{Rapp:2008qc}; evaluating their relation in both weak- 
and strong-coupling limits leads to the pink band in the right panel 
of Fig.~\ref{fig_elec}: it rises with temperature and is indicative for 
a strongly coupled QGP close to $T_c$.
\begin{figure}[!t]
\hspace*{2mm}
\begin{minipage}{0.5\textwidth}
\includegraphics[width=0.92\textwidth]{RAA-v2-elec-tmat.eps}
\end{minipage}
\hspace{-0.2cm}
\begin{minipage}{0.5\textwidth}
\includegraphics[width=0.92\textwidth]{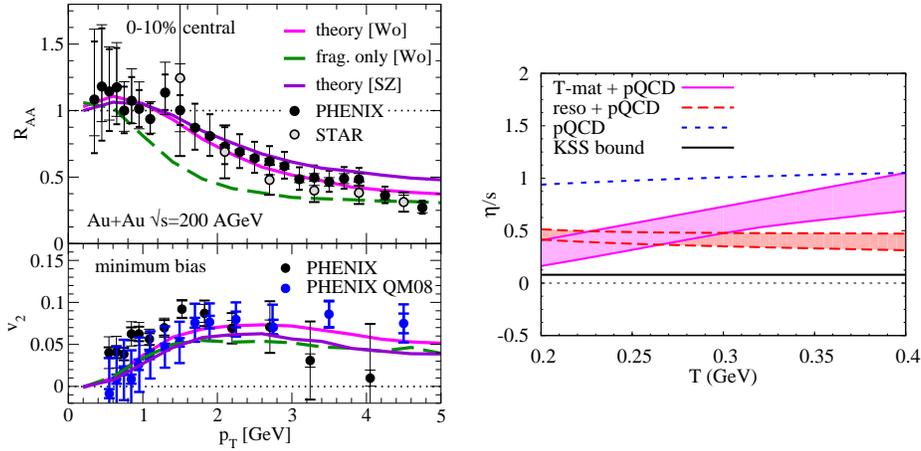}
\end{minipage}
\vspace{-0.35cm}
\caption{Left: $e^\pm$ spectra in Au-Au collisions at RHIC
following from the HQ spectra in Fig.~\ref{fig_hq-spec} after
hadronization and decay~\cite{vanHees:2007me}, compared to
data~\cite{Adare:2006nq,Abelev:2006db,Awes:2008qi}.
Right: estimates of $\eta/s$~\cite{Rapp:2008qc} for pQCD, resonance model 
and $T$-matrix approach.}
\label{fig_elec}
\end{figure}

\section{Conclusions}
\label{sec_concl}
An in-medium $T$-matrix approach (utilizing potentials estimated from 
thermal lattice QCD) has been applied to evaluate HQ interactions in the 
QGP. For charmonia, $S$-wave ground state can survive up to 
$\sim$2$T_c$, roughly in line with lQCD correlators.
For open heavy flavor, resonance-like correlations induce a small 
HQ diffusion coefficient which allows to describe $e^\pm$ data and
suggests a small $\eta/s$ ratio. Several open problems remain, 
e.g., a proper definition of the potential, corrections to the
$T$-matrix approach including radiative ones, and in-medium mass
and width effects.

\vspace{0.5cm}

\noindent
{\bf Acknowledgment}\\
We thank the organizers for the invitation to a very informative workshop.  
This work is supported by a US National Science Foundation CAREER 
award under grant No. PHY-0449489, and by a Bessel Research Award
from the A.-v.-Humboldt foundation.
 

\vfill\eject
\end{document}